\documentclass[final,5p]{elsarticle}

\usepackage{hyperref}
\usepackage{graphicx}
\usepackage{amssymb}
\newcommand{\Tc}{\ensuremath{T_c}}
\newcommand\TBCO{Tl$_2$Ba$_2$CuO$_{6+\delta}$}
\newcommand\YBCO{YBa$_2$Cu$_3$O$_{6+x}$}

\newcommand\Eparac{\ensuremath{E \parallel c}}
\newcommand\dz{\ensuremath{3d_{3z^2-r^2}}}

\journal{Physica C as proceedings for M$^2$S-IX Tokyo}

\begin{document}

\begin{frontmatter}

\title{Possible Microscopic Doping Mechanism in Tl-2201}

\author[Kyoto]{D.\ C.\ Peets\corref{dpeets}}
\ead{dpeets@scphys.kyoto-u.ac.jp}
\author[Waterloo]{D.\ G.\ Hawthorn}
\author[UBC,CIFAR]{Ruixing Liang}
\author[UBC,CIFAR]{D.\ A.\ Bonn\fnref{bonn}}
\ead{bonn@physics.ubc.ca}
\author[UBC,CIFAR]{W.\ N.\ Hardy}

\address[Kyoto]{Department of Physics, Graduate School of Science, Kyoto 
University, Japan}
\address[Waterloo]{Department of Physics \& Astronomy, University of Waterloo, 
ON, Canada}
\address[UBC]{Department of Physics \& Astronomy, University of British 
Columbia, Vancouver, BC, Canada}
\address[CIFAR]{Canadian Institute for Advanced Research, Canada}
\cortext[dpeets]{Corresponding author}
\fntext[bonn]{Requests for samples should be directed to D.A.~Bonn.}

\begin{abstract}

X-ray  absorption  spectroscopy  on  oxygen-annealed,  self-flux-grown
single  crystals of  \TBCO\  suggests a  microscopic doping  mechanism
whereby interstitial  oxygens are  attracted to copper  substituted on
the thallium site, contributing holes  to both the planes and to these
coppers, and  typically promoting  only one hole  to the  plane rather
than two.  These copper substituents  would provide an  intrinsic hole
doping. The evidence for this  is discussed, along with an alternative
interpretation.

\end{abstract}

%
%
%
%

\end{frontmatter}

\section{Introduction}
\label{sec:intro}

Among correlated  electron systems, high-temperature superconductivity
(HTSC) in the cuprates remains the greatest unresolved problem.  Their
anisotropy and line nodes in  the superconducting gap make them highly
susceptible  to  impurities,   thus  accurate  measurements  of  their
electronic properties  rely crucially on  highly perfect, high-quality
single  crystals.   To   maximize  homogeneity  and  fully  understand
experimental results, the carrier doping mechanism must be understood.

On  the less-studied overdoped  side of  the cuprates'  phase diagram,
\TBCO\ (Tl-2201) \cite{Sheng1988} stands out for its particularly flat
CuO$_2$   plane,  a  relatively   simple  crystal   structure  without
(CuO$_2$)$_n$  multilayers  or  CuO  chain layers,  and  for  reaching
$\Tc=0$.     Besides    being     suitable    for    bulk    transport
measurements\cite{Mackenzie1996,Hussey2003},  Tl-2201 has  a non-polar
cleavage  plane  within  its  Tl$_2$O$_2$  double  layer  that  allows
surface-sensitive    single-particle    spectroscopies   to    provide
information  characteristic   of  the  bulk\cite{Plate2005,Peets2007}.
Many    details    of    its    chemical   doping    are    understood
\cite{Shimakawa1993,Wagner1997}, but  not how chemical  doping effects
carrier doping.

In this  article, details of  Tl-2201's doping mechanism  are inferred
from X-ray  absorption spectroscopy  (XAS) measurements on  the copper
$L$  edge.   This technique  measures  the  absorption of  synchrotron
X-rays  as their  energy is  tuned  through the  Cu $2p  - 3d$  atomic
absorption  edge.   These  spectra  are sensitive  to  the  electronic
structure;   because   of  electric   dipole   selection  rules,   the
polarization-dependence   provides  information   on   which  orbitals
participate in each feature.

\section{X-ray Absorption Spectroscopy}\label{sec:xas}

Copper {\slshape L} edge XAS  data were collected at beamline 8.0.1 of
the  Advanced  Light  Source  \cite{Jia1995}, with  a  0.70~eV  energy
resolution, via  total fluorescence yield, and  normalized by incident
intensity, then  against each other  using energy ranges far  from the
edge.   Self-flux  grown,   oxgyen-annealed  Tl-2201  single  crystals
\cite{Peets2007}  were studied,  using  etched square-millimetre-sized
as-grown $ab$-plane faces.  Spectra were collected at room temperature
on  overdoped crystals  with \Tc  s of  9.5, 60  and 69~K  for several
angles between  $E$ and $c$  from 0$^\circ$ to 70$^\circ$,  allowing a
reliable point-by-point extraction of the \Eparac\ component.

\begin{figure}[htb]
\includegraphics[width=\columnwidth]{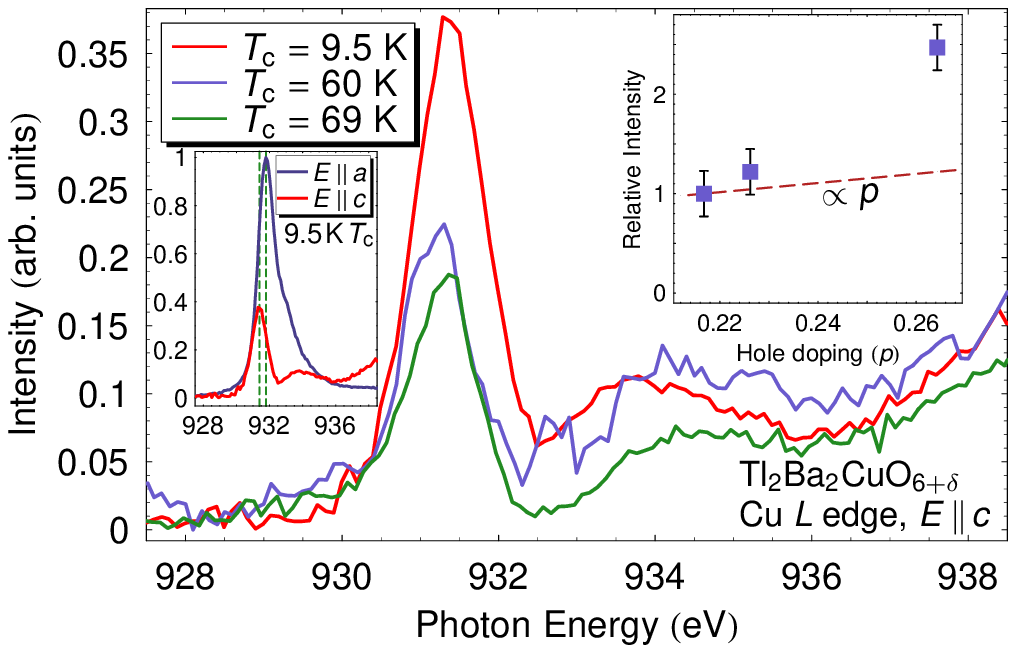}
\caption{\label{fig:edge}Doping  evolution   of  Cu
  {\slshape L$_1$} edge \Eparac\ main peak weight.}
\end{figure}

The  spectra   for  \Eparac\  at  the  $L_1$   edge,  associated  with
excitations  into vacant states  with Cu\dz\  character, are  shown in
Fig.\ \ref{fig:edge}.  The left inset demonstrates that this feature's
energy differs from the in-plane  peak, indicating a different band or
chemical environment.   The right inset shows the  doping evolution of
the  peak's integrated  weight ---  hole dopings  were  extracted from
\Tc\  via  Presland's  phenomenological  formula  \cite{Presland1991}.
This  feature's  intensity changes  far  more  rapidly  than would  be
expected  if  it  were  proportional  to  CuO$_2$-plane  hole  doping,
extrapolating to zero at $p\sim 0.18$, not $p=0$.

It  may be  possible to  attribute the  anomalously steep  increase in
$c$-axis weight with doping to  the copper substituents present at the
4-8\% level in the Tl$_2$O$_2$ double layer.  Cu$^{2+}$ in a Tl$^{3+}$
site would  provide an  intrinsic hole doping,  potentially explaining
the difficulty in underdoping Tl-2201.  An interstitial oxygen atom in
the Tl$_2$O$_2$ double layer could  create one hole oxidizing a nearby
copper substituent  to Cu$^{3+}$ and  contribute only one hole  to the
CuO$_2$  plane.   Occupancy of  this  interstitial  reaches zero  near
optimal doping \cite{Wagner1997},  consistent with the trend observed.
However, since  only a few percent  of thallium sites  are occupied by
copper  and there  are at  most 0.1--0.15  excess oxygens  per formula
unit, the odds of an  oxygen interstitial being adjacent a copper atom
remain low at most dopings.  The strength and doping dependence of the
observed peak may imply  that oxygen dopants preferentially fill sites
adjacent copper substituents.

Traversing    the   overdoped   regime    (changing   $p$    by   0.11
\cite{Presland1991})  requires  changing cation-substituted  Tl-2201's
oxygen     content     by      0.08--0.12     per     formula     unit
\cite{Shimakawa1993,Wagner1997}.  Two  holes per oxygen  would be 0.16
to 0.24 holes per formula unit, leaving 0.05 to 0.13 holes unaccounted
for,  consistent  with   the  concentration  of  copper  substituents.
Previous reports  have linked elevated cation  substitution with lower
\Tc  s and  slower changes  of \Tc\  with oxygen  dopant concentration
\cite{Shimakawa1993,Wagner1997}, but this remained unexplained.

Such  a doping  mechanism would  have several  important implications.
First, ensuring  a homogeneous dopant distribution  may require making
the   {\slshape  cation}   dopants  homogeneous,   significantly  more
difficult  due to  their very  low  mobility.  Second,  if the  oxygen
dopants are randomly distributed after a quench from high temperature,
and  if they  are mobile  at the  storage temperature,  the transition
temperature should  increase over time as oxygens  become trapped near
copper substituents and  supply fewer holes to the  CuO$_2$ plane.  It
may  be  possible  to   take  advantage  of  such  time-dependence  to
controllably change  the doping with  minimal changes to  disorder, as
has been done for underdoped \YBCO\ \cite{Broun2007}.  The addition of
new states  at the  Fermi level may  produce new Fermi  surfaces (thus
far,  only CuO$_2$-plane  bands have  been observed).   Such  a doping
mechanism would  also have  important consequences for  band structure
calculations, in that dopant  oxygens are not randomly distributed and
substituted  copper atoms  must  be considered.   A similar  mechanism
could also be at work in other compounds, including potentially LSCO.

Alternatively,  the  extra \dz\  character  could  originate from  the
CuO$_2$ plane.  Holes  doped into the plane are  shared amongst oxygen
ligands  at  lower  dopings,   to  avoid  the  strong  copper  on-site
repulsion.  If  high ligand hole concentrations produce  a high enough
{\slshape oxygen} on-site  repulsion, some holes may end  up on copper
atoms, with primarily \dz\  character.  This would imply a fundamental
breakdown  in our current  understanding of  the doped  CuO$_2$ plane,
making it a very interesting  possibility.  More work will be required
to    determine   which   picture    is   correct;    observation   of
dopant-clustering or aging, site-specific  studies such as NMR, or the
observation of  an extra  Fermi surface sheet  could all  help clarify
this matter.

\section{Acknowledgements}

The  authors wish  to  thank  J.D.\ Denlinger  and  I.S.\ Elfimov  for
assistance.  This  work was  supported by NSERC,  the CRC  program and
BCSI.  The Advanced Light Source  is supported by the Director, Office
of Science, Office  of Basic Energy Services, of  the US Department of
Energy  under Contract No.\  DE-AC02-05CH11231.

\bibliographystyle{elsarticle-num}
\bibliography{dpeets-M2S}
\end{document}